\documentclass[12pt,preprint]{emulateapj}

\usepackage{natbib}
\newcommand{\bootes}{NDWFS Bo\"otes field}

\slugcomment{}        

\shorttitle{UV ultra-luminous $z\sim3$ LBG}
\shortauthors{F. Bian et al.}

\begin{document}


\title{A UV Ultra-luminous Lyman Break Galaxy at $Z=2.78$ in NDWFS Bo\"otes Field\altaffilmark{1,2,3}} 


\author{Fuyan Bian, Xiaohui Fan, Linhua Jiang}
\affil{Steward Observatory, University of Arizona, 933 N. Cherry Ave. Tucson, AZ, 85721, USA}
\author{Arjun Dey}
\affil{National Optical Astronomy Observatory, 950 North Cherry Avenue, Tucson, AZ 85719, USA}
\author{Richard F. Green}
\affil{Large Binocular Telescope Observatory, University of Arizona, 933 N. Cherry Ave., Tucson, AZ 85721, USA}
\author{Roberto Maiolino}
\affil{INAF Osservatorio Astronomico di Roma
Via di Frascati 33
IT 00040 Monte Porzio Catone
Italy}
\affil{Cavendish Laboratory, Univerisy of Cambridge, 19 J. J. Thomson Ave., Cambridge CB3 0HE, UK}
\author{Fabian Walter}
\affil{Max-Planck-Institut f\"{u}r Astronomie, K\"{o}nigstuhl 17, D-69117 Heidelberg, Germany}
\author{Ian McGreer, Ran Wang}
\affil{Steward Observatory, University of Arizona, 933 N. Cherry Ave. Tucson, AZ, 85721, USA}
\author{Yen-Ting Lin}
\affil{Institute of Astronomy and Astrophysics, Academia Sinica, Taipei, Taiwan ; Institute for the Physics and Mathematics of the Universe, Todai Institutes for Advanced Study, The University of Tokyo, Kashiwa, Chiba, Japan }

\altaffiltext{1}{Based on observations obtained at the Gemini Observatory, which is operated by the Association of Universities for
    Research in Astronomy, Inc., under a cooperative agreement with the
    NSF on behalf of the Gemini partnership: the National Science
    Foundation (United States), the Science and Technology Facilities
    Council (United Kingdom), the National Research Council (Canada),
    CONICYT (Chile), the Australian Research Council (Australia),
    Minist\'{e}rio da Ci\^{e}ncia, Tecnologia e Inova\c{c}\~{a}o (Brazil)
    and Ministerio de Ciencia, Tecnolog\'{i}a e Innovaci\'{o}n Productiva
    (Argentina)}
\altaffiltext{2}{Based on data acquired using the Large Binocular Telescope
(LBT). The LBT is an international collaboration among institutions
in the United States, Italy and Germany. LBT Corporation
partners are: The University of Arizona on behalf of the Arizona
university system; Istituto Nazionale di Astrofisica, Italy; LBT
Beteiligungsgesellschaft, Germany, representing the Max-Planck
Society, the Astrophysical Institute Potsdam, and Heidelberg
University; The Ohio State University, and The Research Corporation,
on behalf of The University of Notre Dame, University of
Minnesota and University of Virginia.}    
\altaffiltext{3}{Based on [in part] data collected at Subaru Telescope, which is operated by the National Astronomical Observatory of Japan. }



\begin{abstract}
We present one of the most ultraviolet (UV) luminous Lyman Break Galaxies (LBGs) (J1432+3358)
at $z=2.78$,
discovered in the NOAO Deep Wide-Field Survey (NDWFS) Bo\"otes field. The $R$-band
magnitude of J1432+3358 is 22.29 AB, more than two magnitudes
brighter than typical $L^{*}$ LBGs at this redshift.
The deep $z$-band image reveals
two components of J1432+3358 separated
by $1.\!{\arcsec}0$ with flux ratio of 3:1.
The high signal-to-noise ratio (S/N) rest-frame UV spectrum shows Ly$\alpha$ emission
line and interstellar medium absorption lines. The absence of \ion{N}{5} and
\ion{C}{4} emission lines, the non-detection in X-ray and
radio wavelengths and mid-infrared (MIR) colors indicate no or weak active galactic nuclei (AGN) ($<10\%$) in this galaxy.
The galaxy shows  broader line profile with the full width half maximum (FWHM) of about 1000~km~s$^{-1}$ and larger 
outflow velocity ($\approx500$~km~s$^{-1}$)  than those of typical $z\sim3$ LBGs. 
The physical properties are derived by fitting the spectral 
energy distribution (SED) 
with stellar synthesis models. 
The dust extinction, $E(B-V)=0.12$,  is similar to that in normal LBGs.
The star formation rates (SFRs) derived 
from the SED fitting and the dust-corrected UV flux are consistent with each other, $\sim$300~M$_{\sun}$~yr$^{-1}$,
and the stellar mass is $(1.3\pm0.3)\times10^{11}$~M$_{\sun}$.
The SFR and stellar mass in J1432+3358 are about 
an order of magnitude higher than those in normal LBGs. 
The SED-fitting results support that J1432+3358 has a continuous star formation history
 with the star formation episode of $6.3\times10^8$~yr.
The morphology of J1432+3358 and its physical properties 
 suggest that J1432+3358 is in an early phase of 3:1 merger process.  
 The unique properties and the low space number density ($\sim$10$^{-7}$~Mpc$^{-3}$)
 are consistent with the interpretation that such galaxies are either found in a short
 unobscured  phase of the star formation or that small fraction of intensive star-forming galaxies 
 are unobscured.
 \end{abstract}




\section{Introduction}
\label{sec:intro}
Over the last decade, the dropout method (the Lyman break technique), 
which uses the fact that little flux is emitted bluewards of the 
Lyman limit (912~\AA), has been fundamental in searching for high-redshift star-forming 
galaxies \citep[e.g.,][]{1996ApJ...462L..17S}.  Spectroscopic follow-up observations show
that the efficiency of this method is high \citep[e.g.,][]{2003ApJ...592..728S,2004ApJ...604..534S}.
Large samples of Lyman Break Galaxies (LBGs) from $z\sim2$ up to $z\sim10$ have been established
\citep[e.g.,][]{2008ApJ...686..230B,2011Natur.469..504B}. These samples of LBGs provide crucial information on
determining the cosmic star formation history \citep[e.g.,][]{1996MNRAS.283.1388M, Lilly:1996fj,Cowie:1996kx}, 
mapping the growth of large scale structures 
\citep[e.g.,][]{1998ApJ...505...18A,Giavalisco:1998uq,Lee:2006qy,Lee:2009qy}, and studying the properties of dark matter halos hosting
the LBGs. 

Optical and near-infrared (NIR) photometric and spectroscopic observations 
of these galaxies reveal the properties of the UV-selected galaxies at $z\sim2-3$. 
The median stellar mass of the $z\sim3$ LBGs is about $2.4\times10^{10}$~M$_{\sun}$ \citep[e.g.,][]{2001ApJ...562...95S},
and the mean star formation rate (SFR) derived from the H$\alpha$ and UV luminosity is
about 30 M$_{\sun}$~yr$^{-1}$ \citep[e.g.,][]{2006ApJ...647..128E}. 
The median dust extinction ($E(B-V)$) is around 0.15 \citep[e.g.,][]{2001ApJ...562...95S}.
The $z\sim2-3$ LBGs show compact morphologies (half-light radii, $r_e<0.\!\arcsec5$)
in Hubble space telescope (HST) images \citep[e.g.,][]{1998hdf..symp..121G}.

However, most of optical/NIR surveys for high-redshift galaxies are 
deep field surveys with survey area less than 1~deg$^2$.  So far, the largest $z\sim2-3$ LBGs 
survey with spectroscopic redshifts only covers a total area of 
around one deg$^2$ with $>2000$ spectroscopic 
redshifts \citep[e.g.,][]{2003ApJ...592..728S,2009ApJ...692..778R}. 
Due to the small survey volume, combined with the rapid decline of galaxy luminosity function at the bright end, 
these surveys are not suited to reveal the most luminous and most massive systems;
previous studies have been  focusing on
the LBGs with luminosity of $L^{*}$ or sub-$L^{*}$ ($r>24.5$).
A sample of bright LBGs was discovered in Sloan Digital Sky Survey (SDSS), 
however, HST follow-up observations show that these galaxies are unresolved point
sources indicating that these objects are quasars rather than galaxies \citep{Bentz:2008yq}.
To date only one unlensed LBG at $z\sim3$ with $R$-band magnitude brighter
than 22.5 has been found \citep{2008ApJ...681L..57C}. 
The nature and properties of this type of galaxy are still unknown, so 
it is important to build up a sample of these UV ultra-luminous galaxies
and preform detailed follow-up observations on them.

Finding UV ultra-luminous $z\sim2-3$ galaxies requires wide-field surveys
with deep, multi-color broad band images.
In this paper, we report a UV ultra-luminous LBG, J1432+3358, 
with $R_{\rm AB} = 22.29$
at $z=2.78$ discovered in 
the NOAO Deep Wide Field Survey (NDWFS) Bo\"otes field. Through this paper,
we adopt $\Omega_m = 0.3$, $\Omega_\Lambda=0.7$, and $H = 70$~km~s$^{-1}$~Mpc$^{-1}$ \citep{2007ApJS..170..377S}.
All the magnitudes are AB magnitudes.

\section{Observations}

\label{sec:obs}
In 2008 and 2009, we carried out  deep $U$- and $Y$-band imaging of 
the 9~deg$^2$ NOAO Deep Wide-Field Survey (NDWFS) Bo\"otes 
Field  \citep{1999ASPC..191..111J}.  Our survey used the $2\times8.4$~m Large 
Binocular Telescope (LBT) \citep{2010SPIE.7733E..10H} equipped with 
two prime focus Large Binocular Cameras  \citep[LBC,][]{2008A&A...482..349G}.
This new LBT survey builds on the available unique multi-wavelength data of the {\bootes}
and fills in two critical wavelength gaps at 3500~{\AA} and 1$\mu$m
with the $U$ and $Y$ bands.
The deep $U$-band images (25.2 AB magnitude with 5$\sigma$ detection), together with the existing 
$B_{\rm W}$- and $R$-band images
taken with the Mosaic CCD camera on Kitt Peak 4~m Mayall telescope, allow us to 
search for the star-forming galaxies at $z\sim3$ using the U-dropout technique, and total
15,000 LBG candidates   are selected based on the $U-B_{\rm W}$ and $B_{\rm W}-R$ color-color diagram,
with the selection criterion being
\begin{eqnarray}
 U-B_{\rm W} > 1.0, \nonumber \\
 B_{\rm W}-R<1.9, \nonumber \\
 B_{\rm W}-R < U-B_{\rm W}-0.1, \nonumber \\
 R<25.0 .\label{z3selection}
\end{eqnarray} 
The typical image quality in the $R$-band is 1$''$ and thus can not resolve typical LBGs. Nevertheless, 
the large survey area allows us to select and study the most UV luminous LBGs at $z\sim3$ with $R<22.5$ ($L>7L^*$ at $z\sim3$).

Spectroscopic follow-up observations of 12 of bright LBG candidates were obtained using the blue channel spectrograph
on 6.5~m Multiple Mirror Telescope (MMT) on 2010 April 15.  Typically 20-40~minutes exposures were taken for each candidate. The wavelength
coverage is 4000-7500~\AA.
One out of 12 candidates was confirmed as a UV ultra-luminous LBG (J1432+3358) at $z=2.78$, and the 
coordinates of this galaxy are R.A.=14$^{\rm h}$32$^{\rm m}$21$^{\rm s}$.84 and Decl.=33$^{\rm o}$58$'$18$''$.2, J2000.
The remaining eleven candidates are all quasars in the redshift range of $2<z<3$. All these quasars 
show broad Ly$\alpha$, \ion{N}{5}, and \ion{C} {4} emission lines and they are all point sources.

A high signal-to-noise ratio (S/N) spectrum of J1432+3358 was obtained 
with the 8.2~m Gemini-N telescope and GMOS instrument on 2011 March 9 and 10 (Program ID: GN-2011-C-5).
The sky was clear and the resulting image quality was $0.\!\arcsec6-0.\!\arcsec7$.
The total exposure time was 4~hr and was divided into eight 30-min individual exposures. 
The slit width was 1$''$. The B600-G5307 grating was used, and two central wavelengths
of 5200~{\AA} and 5300~{\AA} were used to fill the gaps between different CCD chips. 
The wavelength coverage was from 4000~{\AA} to 6500~{\AA}, and the spectral resolution ($R=\lambda/\delta\lambda$)
is 850. The airmass of the object during the observing was about 1.05, thus we did not use the parallactic 
angle. The slit oriented in P.A.=-60 degrees (300 degrees), which was roughly along the galaxy extended
direction (Figure~1). The spectrophotometric standard G191-B2B was observed for flux calibration, and 
a CuAr arc lamp was used for wavelength calibration.
The spectra were reduced and calibrated using standard Gemini IRAF
package. The final spectrum
has been smoothed by 4~{\AA}. The S/N per spectral element ($\sim4$~{\AA}) is 8-10.

MIR photometry of J1432+3358 was obtained by the Spitzer Deep Wide-Field Survey \citep[][]{2009ApJ...701..428A}.
In addition, $H$-band image with one hour exposure was obtained using the SWIRC on 6.5~m MMT on 2012 Janurary 5.
We also got deep z-band image from the Subaru B\"ootes field survey (B\"ootesZ Survey) (Y. Lin et al. 2012, in preparation). 
The image quality of the $z$-band image is 
$0.\!\arcsec5$. 
The total magnitudes of J1432+3358 are measured with SExtractor \citep{1996A&AS..117..393B},
and are listed in table~\ref{morphology}. 

\section{Results}
\label{sec:results}

\subsection{Lensed or Unlensed?}

In the last decade, a sample of bright lensed high-redshift 
galaxies has been established through systematic searches
towards galaxy clusters \citep[e.g.][]{Mehlert:2001fk,Sand:2005fk} and red galaxies 
in the SDSS images \citep[e.g.][]{Smail:2007qy,Allam:2007qy,Diehl:2009lr,Lin:2009vn}. 
It is crucial to determine whether J1432+3358 is lensed or not.
Studies suggest that the total fraction of high-redshift
galaxies and quasars that are lensed is small \citep[e.g.,][]{Turner:1984lr,Jain:2011lr}, 
but the lensing contribution becomes larger with increasing brightness.
\citet{Jain:2011lr} find that 
the lensing contribution becomes significant when $L>10L^*$,
and about 1/3-1/2 of the LBGs with $L=7L^*$ are lensed galaxies 
\citep{van-der-Burg:2010lr}.

\begin{figure*}
\center
\includegraphics[scale=1.1]{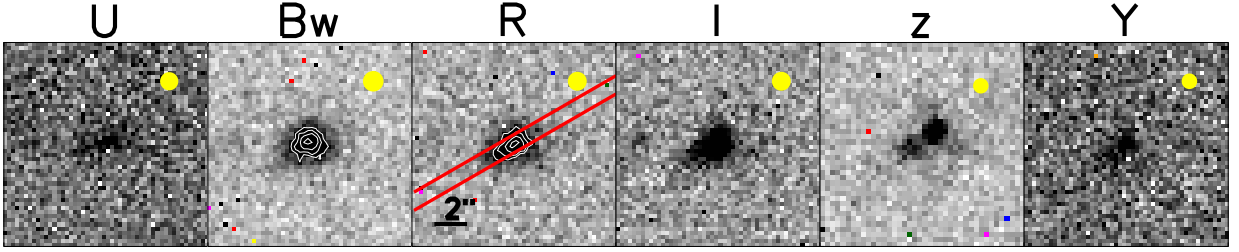}
\caption{The $U$-, $B_{\rm W}$-, $R$-, $I$-, $z$-, and $Y$-band images of J1432+3358. 
A contour plot is also shown in $B_{\rm W}$- and $R$-band images.  
The yellow filled circles represent the size of point spread functions (PSF) in each image.
The typical image qualities for  $U$-, $B_{\rm W}$-, $R$-, $I$-, $z$-, and $Y$-band images
are $1.\!\arcsec0$, $1.\!\arcsec3$, $1.\!\arcsec2$, $1.\!\arcsec2$, $0.\!\arcsec5$ and $0.\!\arcsec6$, respectively. 
The slit position and orientation 
are also shown in $R$-band image.
\label{img}}
\end{figure*}

The structure of J1432+3358 is well resolved by the ground-based imaging 
observations.
The broad band $U$-, $B_{\rm W}$-,  $R$-, $I$-, $z$- and $Y$-band images (Figure~\ref{img}) of J1432+3358 
show extended morphology.
The deep multi-band images show that there is no foreground lensing galaxy
and the morphology of the galaxy is also consistent with being
unlensed. Especially, in the deep Subaru $z$-band image with image quality of $0.\!\arcsec5$,
J1432+3356 is revolved into two 
components, and separation of these two parts is by $1.\!\arcsec0$ (7.8~kpc). 
These two components do not show stretched arc structures at the resolution of the image. 
Furthermore, 
the central wavelengths of Ly$\alpha$ emission from these two
components have a small offset ($237\pm23$~km~s$^{-1}$)  and the Ly$\alpha$ flux ratio of these two components 
is not consistent with the continuum flux ratio of these two components
(see detail in section 3.3).
Meanwhile, the spectrum of J1432+3358 does not show any other redshift systems that could be
from the lensing galaxy.
Therefore, we conclude that J1432+3358 is not a lensed galaxy.

\begin{table*}
\begin{center}
\caption{Magnitude and morphological properties of J1432+3358.\label{morphology}}
\begin{tabular}{ccccccc}
\tableline
\tableline
Filter & Magnitude\tablenotemark{a}&$r_e$\tablenotemark{b}&$n$\tablenotemark{c}&$b/a$\tablenotemark{d}&$\theta$\tablenotemark{e}&$\chi^2/\nu$\tablenotemark{f}\\
\tableline
 $U$&$24.35\pm0.13$&-&-&-&-&-\\
$B_{\rm W}$&$23.26\pm0.02$&$0.73\pm0.20$&$3.10\pm0.77$&$0.49\pm0.05$&$-73.17\pm3.75$&1.310\\
$R$&$22.29\pm0.03$&$0.89\pm0.04$&$1.23\pm0.27$&$0.45\pm0.04$&$-63.81\pm2.84$&1.078\\
$I$&$22.20\pm0.03$&$0.82\pm0.05$&$1.37\pm0.37$&$0.38\pm0.05$&$-55.04\pm3.05$&1.104\\
$z$&$22.13\pm0.03$\\
$z(a)$\tablenotemark{g}&$22.62\pm0.04$\tablenotemark{i}&$0.26\pm0.03$&4&$0.69\pm0.08$&$14.04\pm13.21$&1.168\\
$z(b)$\tablenotemark{h}&$23.71\pm0.07$\tablenotemark{i}&$0.21\pm0.03$&1&$0.67\pm0.14$&$82.43\pm19.12$&1.168\\
$Y$&$22.19\pm0.09$&-&-&-&-&-\\
$H$&$21.54\pm0.36$&-&-&-&-&-\\
IRAC1&$20.88\pm0.06$&-&-&-&-&-\\
IRAC2&$20.66\pm0.07$&-&-&-&-&-\\
IRAC3&$20.69\pm0.37$&-&-&-&-&-\\
IRAC4&$20.42\pm0.33$&-&-&-&-&-\\
\tableline
\end{tabular}
\tablenotetext{1}{Total AB magnitude from SExtractror.}
\tablenotetext{2}{Effective radius ($''$) from the GALFIT fitting.}
\tablenotetext{3}{S\'ersic index from the GALFIT fitting.}
\tablenotetext{4}{The ratio of minor axis (b) and major axis (a) from the GALFIT fitting.}
\tablenotetext{5}{The position angle from the GALFIT fitting.}
\tablenotetext{6}{The reduced chi-square from the GALFIT fitting.}
\tablenotetext{7}{The GALFIT fitting results of the brighter components in $z$-band image. }
\tablenotetext{8}{The GALFIT fitting results of the fainter components in $z$-band image. }
\tablenotetext{9}{The magnitude is from the GALFIT fitting.}
\end{center}
\end{table*}

\subsection{Morphology}

The deep $z$-band image reveals two components in J1432+3358.
We use GALFIT  \citep{2002AJ....124..266P} to fit the light distribution of 
these two components with
either exponential disk or DeVaucouleurs profiles.
We find that the DeVaucouleurs profile 
is better to fit the brighter component and 
the exponential disk is better to fit the fainter component.
The effective radius are $0.\!\arcsec26\pm0.\!\arcsec03$ and $0.\!\arcsec21\pm0.\!\arcsec03$, respectively, 
which correspond to about 2.0~kpc in physical size. 
This size is comparable
to the typical size of LBGs at $z\sim3$ \citep[e.g.][]{Ferguson:2004lr}. 
The distance between these two components is about $1.\!\!\arcsec0$, which is
about 7.8~kpc.
The brightness ratio between these two components in J1432+3358 is 3 to 1.
Assuming a simple relation between luminosity and stellar mass, the mass ratio
of the two components is also 3 to 1 implying that it is a 3:1 merger. These two components are
barely resolved by the $z$-band image. 
Further high resolution HST follow-up imaging observations will help us 
to fit the systems more accurately and  
to reveal more detailed structure of this galaxy.

The morphology of J1432+3358 in $B_{\rm W}$-band image and that in $R$- and $I$-band images
look different (figure~1).  
To characterize the morphology in these bands,  we use
GALFIT to model the galaxy light distribution in the $B_{\rm W}$-, $R$-, 
and $I$-band images of J1432+3358. 
The S{\'e}rsic profile \citep{1963BAAA....6...41S} is used 
to fit the light profile of J1432+3358. The GALFIT fitting results are listed in table~\ref{morphology}.
The $R$- and $I$-band morphologies of J1432+3358 can be 
well fit by a disk-like profile with S\'ersic index $n\approx 1.3$,
while the S{\'e}rsic index is about 3.10 for the morphology in 
$B_{\rm W}$-band image,
which is much larger than that in $R$- and $I$-band images (Table 1).
Furthermore, the galaxy light distribution in $B_{\rm W}$-band image can not be well fitted by a single 
S{\'e}rsic profile component, 
with reduced chi-square of 1.310. The $B_{\rm W}$-band morphology is round rather than elongated. 
The strong Ly$\alpha$ emission
line lies within the $B_{\rm W}$ filter at the redshift of 2.78.
 We therefore interpret the diffuse $B_{\rm W}$ morphology as being likely due to a diffuse Ly$\alpha$ 
 halo around J1432+3358 \citep[c.f.,][]{2011ApJ...736..160S}.  The Ly$\alpha$ photons 
 from the central galaxy are scattered by the neutral hydrogen gas in the galaxyÕs circum-galactic medium (CGM)
 and form the diffuse Ly$\alpha$ halo.
 We do not carry out the fitting on $U$- and $Y$- band images, because their
S/N is too low.


\begin{figure}
\center
\includegraphics[scale=0.45]{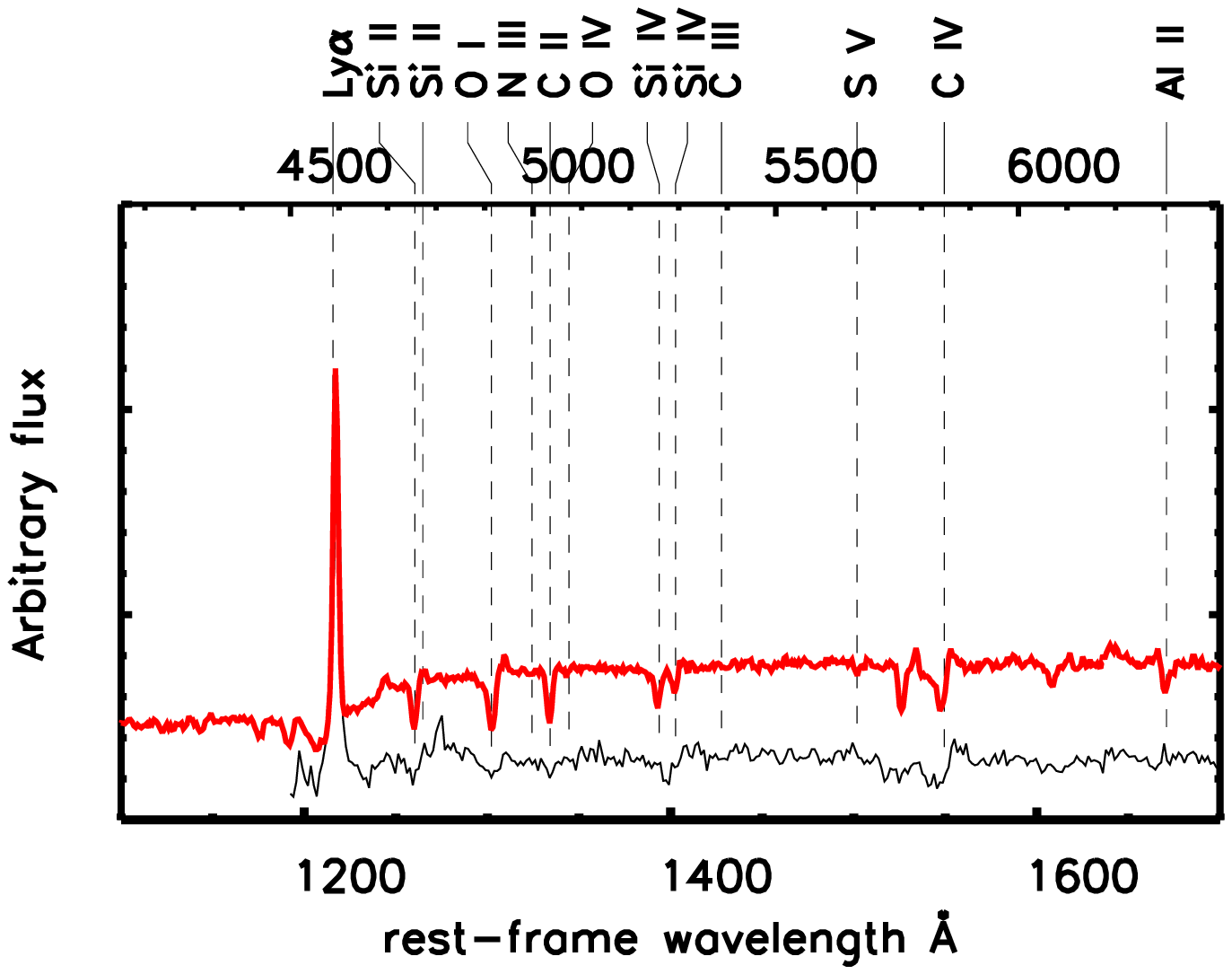}
\caption{The rest-frame UV spectrum of J1432+3358 in the wavelength range 1100-1700~\AA (the thin black solid
curve). For comparison, the composite spectrum of a sample of $z\sim3$ LBGs is also shown
with a red thick solid curve \citep{2003ApJ...588...65S}. The top x-axis represents the observed-frame wavelength.
Both spectra are scaled by the peak value of the Ly$\alpha$ emission line, and the composite spectrum
is shifted by +0.1 in flux density direction for clarity.
\label{spec}}
\end{figure}

 \begin{table*}
\begin{center}
\caption{Ly$\alpha$ emission and strong interstellar absorption lines.\label{line}}
\begin{tabular}{ccccccc}
\tableline
\tableline
line & $\lambda_{\rm rest}$\tablenotemark{a} 
&$\lambda_{\rm obs}$  &EW&FWHM& redshift & $\Delta v\tablenotemark{b}$ \\
&\AA&\AA&\AA&km~s$^{-1}$&&km~s$^{-1}$\\
\tableline
 Ly$\alpha$\tablenotemark{c}& 1215.08&$ 4592.99\pm 0.59$&$   34.97\pm 2.74$&$  1431\pm   81$&$   2.780\pm 0.000$&$  392\pm  38$\\
 Ly$\alpha(a)$\tablenotemark{d}& 1215.08&4597.41\tablenotemark{g}&$   27.46\pm 2.59$&$   852\pm   56$&$   2.784$&$  680$\\
 Ly$\alpha(b)$\tablenotemark{e}& 1215.08&4582.52\tablenotemark{g}&$    9.86\pm 1.38$&$  1073\pm  151$&$   2.771$&$ -291$\\
  Ly$\alpha(c)$\tablenotemark{f}&1215.08 &$4601.05\pm0.36$&$4.5\pm1.5$&$200\pm25$&2.787&$923\pm25$\\
   \ion{Si}{2}& 1259.83&$ 4752.42\pm 1.83$&$   -1.89\pm 0.52$&$  1004\pm  288$&$   2.772\pm 0.001$&$ -222\pm 115$\\
   \ion{O}{1}& 1302.69&$ 4912.05\pm 4.33$&$   -2.65\pm 0.89$&$  2031\pm  643$&$   2.771\pm 0.003$&$ -346\pm 264$\\
    \ion{Si}{4}& 1393.18&$ 5274.68\pm 2.00$&$   -2.11\pm 0.50$&$  1041\pm  259$&$   2.786\pm 0.001$&$  872\pm 113$\\
    \ion{C}{4}& 1548.91&$ 5825.82\pm 2.41$&$   -4.04\pm 0.83$&$  1901\pm  384$&$   2.761\pm 0.002$&$-1101\pm 123$\\
\tableline
\end{tabular}
\tablenotetext{1}{Air wavelengths.}
\tablenotetext{2}{The line velocity relative to the systemic redshift. The negative (positive) values correspond to blueshift (redshift).} 
\tablenotetext{3}{The whole Ly$\alpha$ emission.}
\tablenotetext{4}{The primary peak of the Ly$\alpha$ emission corresponding to the `a' component in Figure~2. }
\tablenotetext{5}{The secondary peak of  the Ly$\alpha$ emission corresponding to the `b' component in Figure~2.}
\tablenotetext{6}{The Ly$\alpha$ emission from the faint component corresponding to the `c' component in Figure~2. The information of 
this component is derived from Extractor measurement and Gaussian fitting in the 2D spectra image.}
\tablenotetext{7}{To deblend the two peaks of the Ly$\alpha$ emission line,  the central wavelength values are set to the 
peak values of these two lines and fixed during the fitting.}
\end{center}
\end{table*}

 \begin{figure}
\center
\includegraphics[scale=0.5]{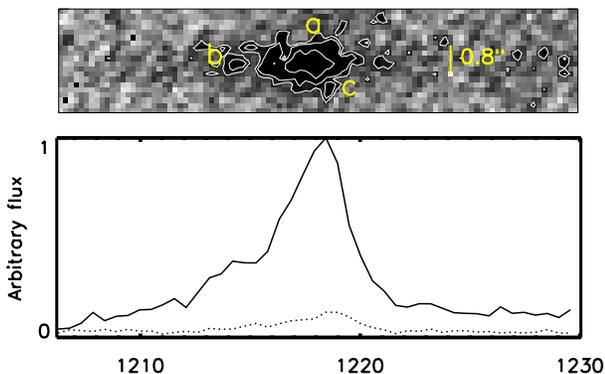}
\caption{The 2D spectrum (top panel) and the 1D spectrum (top panel) of the Ly$\alpha$ emission line.
There are 3 significant components resolved in 2D spectrum image. The components of `a' and `b'
correspond to the redshifted stronger peak and the blueshifted weaker peak, respectively. The 
component of `c' corresponds to the Ly$\alpha$ emission from the fainter component which was
resolved in $z$-band image. The central wavelength difference between components `a' and `c'
is about 3.6~\AA.  
\label{lya}}
\end{figure}

\begin{figure}
\center
\includegraphics[scale=0.5]{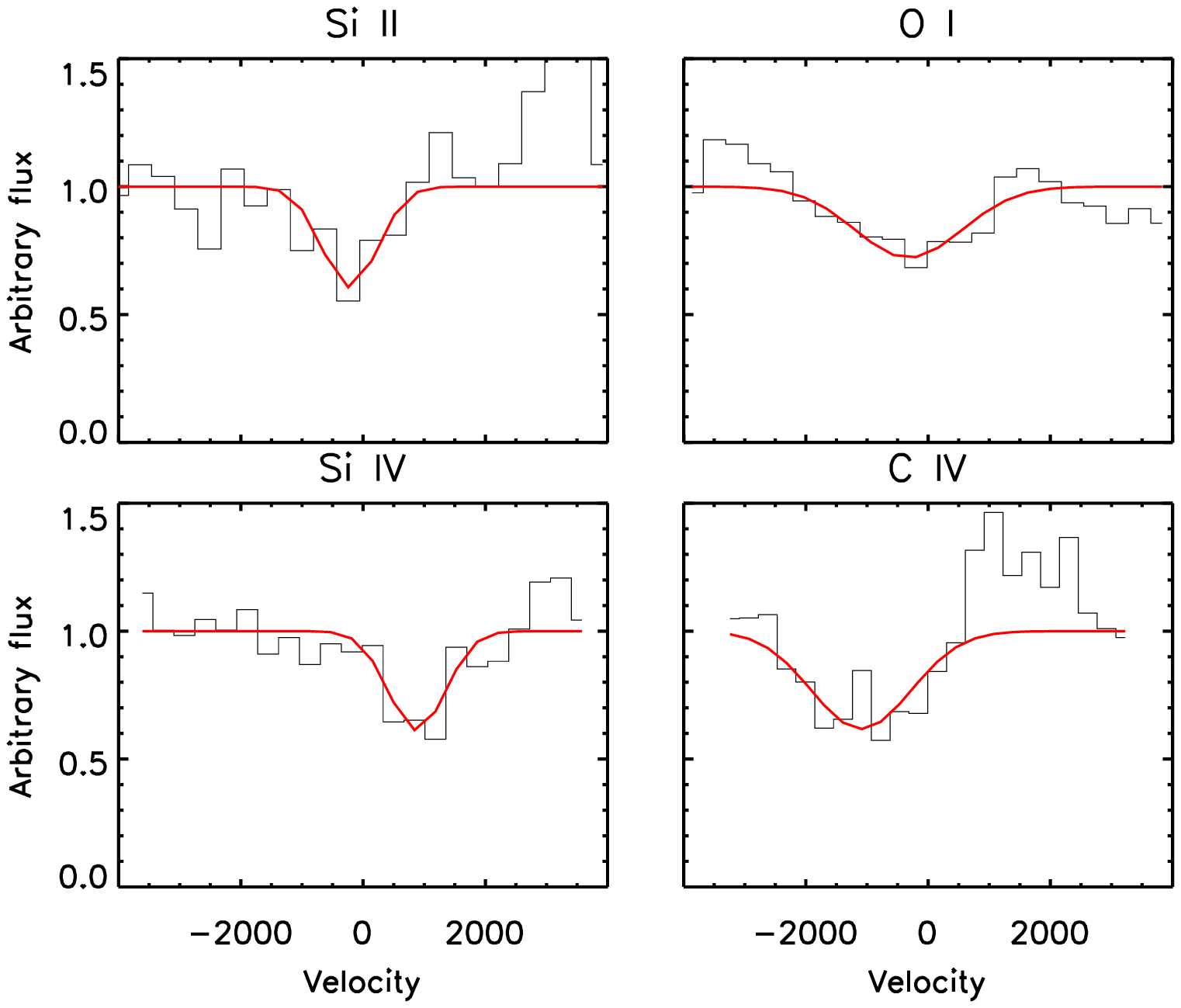}
\caption{The spectra around the absorption lines (the back curves in histogram mode) and best-fit Gaussian models
(the red curves). The 
flux is scaled by the continuum flux. \label{spec_abs}}
\end{figure}

\subsection{Spectroscopy}
Figure~\ref{spec} shows the high S/N rest-frame UV spectrum of J1432+3358 taken using Gemini GMOS-N.
The spectrum covers the rest-frame wavelength from 1100-1700~{\AA}, and shows the strong
Ly$\alpha$ emission line and a few absorption features from the interstellar medium (ISM)
(e.g., \ion{Si}{2}, \ion{O}{1}, \ion{Si}{4}, and \ion{C}{4}).  In the spectrum, there is no prominent \ion{N}{5} 1240 emission line detected
at the level of equivalent width of 2~{\AA} ($1\sigma$ level) corresponding to  
rest-frame equivalent width (EW$_0$) of 0.5~{\AA}, which is much smaller than the EW$_0$ of 
$18\pm10$~{\AA} measured in a sample of bright quasars \citep[][]{Forster:2001ly}.
Similarly, the \ion{C}{4} 1548, 1551 doublet line shows an
absorption rather than emission feature.

J1432+3358 is not detected at the X-ray energies (0.5-7.0~keV) 
with 30~ks {\it Chandra} observation (P.I. Murray Obs ID 13134)
\citep[also see,][]{Murray:2005ly}. The flux limit of the X-ray data 
is about $4\times10^{-16}$~erg~cm$^{-2}$~s$^{-1}$. 
By assuming the X-ray spectrum as a common AGN 
power-law spectrum with photon index $\Gamma=1.7$ \citep[e.g.,][]{Kenter:2005gf}, the 
Galactic neutral hydrogen column density $\rm N_{\rm H} = 1\times10^{20}$~cm$^{-2}$ \citep[e.g.,][]{Dickey:1990fk},
and the X-ray-to-optical power-law slope $\alpha_{\rm ox}=-1.45$ \citep[e.g.,][]{Just:2007lr},
we find that less than 10\% of the optical radiation could be from the AGN. 
J1432+3358 is not detected at 20cm in radio wavelength \citep{2002AJ....123.1784D}, 
with $1\sigma$ sensitivity limit of 28$\mu$Jy corresponding
to a specific luminosity of $4.7\times10^{30}$~erg~s$^{-1}$~Hz$^{-1}$.
This radio limit can not put a good constraint on the AGN activity in J1432+3358 by adopting  the 
relation between optical and radio luminosity derived from the SDSS quasar sample
\citep{White:2007uq}. 
The rest-frame NIR excess does not show in the {\it Spitzer} bands, which suggests absence 
of hot dust and obscured AGN.
The IRAC [3.6]-[4.5] and [5.8]-[8.0] colors are located out of the AGN selection area
in MIR color-color space \citep[e.g,][]{Lacy:2004fk,Stern:2005lr}.
The rest-frame UV spectrum, the non-detection in X-ray and radio bands and MIR colors 
imply that there is no or only weak AGN ($<10\%$) in J1432+3358, therefore, 
we suggest that the UV emission is dominated by the emission from massive stars rather than
a central AGN.

The properties of the well detected (S/N$> 3$) lines in the rest-frame UV spectrum 
are listed in table~\ref{line}. 
Because the observed wavelength of Ly$\alpha$ emission and ISM absorption 
lines are affected by galactic-scale outflows, the redshifts from these lines
do not represent the systemic redshift \citep[e.g.,][]{2005ApJ...629..636A} . Typically, the systemic 
redshift can be derived from absorption lines that are clearly associated with
photospheric features (e.g., \ion{S}{5} 1502, \ion{C}{3} 1176, \ion{O}{6} 1343, etc)
\citep[e.g.,][]{Dey:1997mz,2003ApJ...588...65S}, but we can not
identify these lines from the spectrum at current S/N level.
Therefore, we follow \citet{2005ApJ...629..636A}, in which they calibrate 
the redshifts of the Ly$\alpha$ and ISM absorption lines with the H$\alpha$ nebular line in a sample 
of $z\sim2-3$ LBGs \citep[see details in][]{2005ApJ...629..636A}. The scattering of this relation is less than 0.0015.
Here the systemic redshift of J1432+3358 is estimated from the Ly$\alpha$ and absorption lines 
using the equations 1 and 2 in \citet{2010ApJ...717..289S}, which are 2.7745 and 2.7755, respectively.
This result is also consistent with the systemic redshift calculated using the equation 3 in \citet{2005ApJ...629..636A}.
The systemic redshift of $2.775\pm0.001$ is adopted for further analysis in this paper.

The composite spectrum of LBGs at $z=2-3$ is also shown in Figure~\ref{spec} for 
comparison  \citep{2003ApJ...588...65S}. The continuum shape of the spectrum is similar to that of LBG composite
spectrum, indicating that the dust extinction in J1432+3358 is similar to the dust 
extinction in the typical LBGs. This property is also supported by the SED fitting 
results (see details in section~\ref{sec:physical}).
 
The prominent  Ly$\alpha$ emission line is shown in Figures~\ref{lya} and \ref{spec} 
with EW$_0$ of 35~\AA. 
In Figures~\ref{lya}, 
the Ly$\alpha$ emission shows two peaks separated by 970~km~s$^{-1}$ and the full width at half maximum (FWHM)
of the Ly$\alpha$ emission line is  $\sim1500$~km~s$^{-1}$,  which is much larger than the  
FWHM(Ly$\alpha$) = $450\pm150$~km~s$^{-1}$ typically found in $z\sim3$ LBGs \citep{2003ApJ...588...65S}.
The secondary peak is real rather than noise, because 
the double-hump profile is exhibited in the individual spectra of the eight individual 30-min exposures.
Two Gaussian profiles are used to deblend the
Ly$\alpha$ emission lines. The FWHMs of the two Ly$\alpha$ components are $\sim1000$~km~s$^{-1}$.
Like most of the LBGs with multiply-peaked Ly$\alpha$ emission \citep{Kulas:2012lr},
J1432+3358 shows stronger redshifted Ly$\alpha$ emission than blueshifted.
The separation of the two Ly$\alpha$ peaks is $\sim1000$~km~s$^{-1}$, which is comparable to 
the separation in other multiple-peaked LBGs  \citep{Kulas:2012lr}. Such a
Ly$\alpha$ line profile is observed in the expansion shell model \citep[]{2006A&A...460..397V},
which predicts that a primary Ly$\alpha$ peak is redshifted by approximately two times
the expansion velocity and another blueshifted Ly$\alpha$ emission is located around the expansion
velocity.

In the 2-dimensional (2D) spectral image of the Ly$\alpha$ emission (Figure~\ref{lya}), 
there are three significant components of Ly$\alpha$ emission
detected. The `a' and `b' components which are resolved in the wavelength direction
correspond to the double-peak feature
in the 1-dimensional (1D) spectrum.  The `c' component resolved in the
slit (spatial) direction is the Ly$\alpha$ emission from the fainter component
detected in the $z$-band image.  To determine the wavelength of `c' component,
SExtractor and GALFIT are used to obtain the centroid and GAUSSIAN fitting central 
positions of `a' and `c' components.
The offsets of the central wavelength
between `a' and `c' components are 3.90~{\AA} and 3.25~{\AA} from the measurements with above two methods, 
respectively. We adopt the $3.64\pm0.36$ as the offset of the central wavelength between `a' and `c' components,
which corresponds to a velocity difference
of $237\pm23$~km~s$^{-1}$. 
The Ly$\alpha$ flux ratio between the components `a' and `c' is about 7:1 from the SExtractor measurement,
which is higher than
the continuum flux ratio between these two components. The velocity offset and the difference in flux ratio of Ly$\alpha$
and continuum in components of `a' and `c' also support that J1432+3358 is not a lensed system.


The spectrum also shows weak absorption lines originated from the interstellar median (ISM), 
which are \ion{Si}{2}, \ion{O}{1}, \ion{Si}{4}, and \ion{C}{4} lines (Figure~\ref{spec_abs}). The EW$_0$ of these absorption
features are a few \AA, comparable to those in typical LBGs.  However, the FWHMs
of these lines are 1000~km~s$^{-1}$ or even larger (Table~\ref{line}),
about two times larger than those in typical LBGs. 
The absorption lines 
show blueshifts with velocities of 200-1000~km~s$^{-1}$,  which can be 
interpreted by a galactic scale outflow model \citep{2000ApJS..129..493H,2010ApJ...717..289S}. 
The average outflows velocity estimated from \ion{Si}{2}, \ion{O}{1} and \ion{C}{4} absorption lines is
$-556\pm103$~km~s$^{-1}$. We do not use the \ion{Si}{4} 1392 absorption line for the analysis,
because the measurements of this line is contaminated by the \ion{S}{4} 1402
\citet{2010ApJ...717..289S} find that average outflow velocity of LBGs derived from the ISM absorption
 lines is $164\pm16$~km~s$^{-1}$ with a wide range distribution from 0 to 500~km~s$^{1}$, and we find 
the outflow velocities in J1432+3358 are larger than the outflow velocities in most ($>98\%$) of LBGs.
.

\begin{figure}
\center
\includegraphics[scale=0.5]{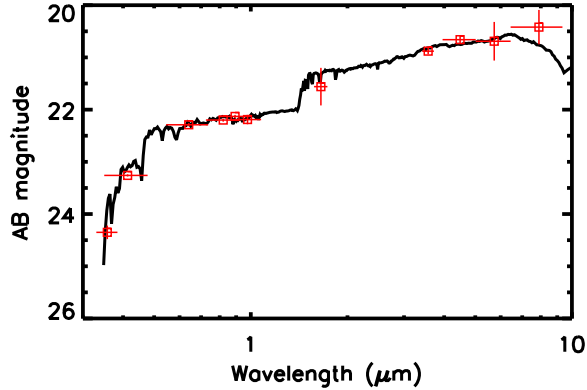}
\caption{The best-fit stellar synthesis model (solid curve) 
and photometric data points from $U$-band to IRAC4-band (red squares). \label{sed}}
\end{figure}

\subsection{Physical Properties}
\label{sec:physical}
An IDL-based code FAST (Fitting and Assessment of Synthetic Templates) \citep{2009ApJ...700..221K} 
is used
to fit the broad band ($U$, $B_{\rm W}$, $R$, $I$, $Y$, $H$, IRAC1, IRAC2, IRAC3, and IRAC4)
photometry with stellar population synthesis models \citep[BC03]{2003MNRAS.344.1000B} and 
to derive the physical properties of J1432+3358.
First, we use an exponential star formation rate (SFR$\propto\exp{(-t_{sf}/\tau)}$) with a
Salpeter initial mass function (IMF) \citep{1955ApJ...121..161S} and solar metallicity.
We adopt the dust extinction law proposed by \citet{Calzetti:2000vn} and the intergalactic medium (IGM)
absorption model in \citet{Madau:1995lr}.
 From the SED fitting,
we place constraints on the exponential star formation time scale, $\log(\tau(\rm yr))=10.00^{+1.00}_{-1.66}$.
The large $\tau$ value indicates that the SFR declines slowly, suggesting that J1432+3358 has 
continuous constant star formation history.
Therefore, we adopt a constant
star formation history model for further analysis.  In Figure~\ref{sed}, we show the best-fit
spectral energy distribution overlaid on the optical ($U$, $B_{\rm W}$, $R$, and $I$, Y), NIR ($H$),  and MIR
(IRAC1 (3.6$\mu$m), 2 (4.5$\mu$m), 3 (5.8$\mu$m), 4 (8.0$\mu$m)) photometry. 
From the fitting, we find that the galaxy age, $\log(t_{sf} (\rm yr))=8.8^{+0.2}_{-0.2}$,
the SFR is $280^{+70}_{-60}$~M$_{\sun}$~yr$^{-1}$, 
the dust extinction is 
$E(B-V) = 0.12\pm0.02$, and the stellar mass is $(1.3\pm0.3)\times10^{11}$~M$_{\sun}$.
The SFR derived from the dust-corrected UV luminosity is 310~M$_{\sun}$~yr$^{-1}$ \citep{1998ARA&A..36..189K},
which is consistent with that derived from the SED fitting. The SFR in J1432+3358
thus is one order of magnitude higher than that in typical $z\sim3$ LBGs \citep[]{2001ApJ...562...95S} .
The SED fitting suggests that the star formation age ($t_{sf}$) is about
two times longer than the median star formation time in typical $z\sim3$ 
LBG sample \citep{2001ApJ...562...95S}. This age is also longer than 
the typical time scale of the starburst in submillimeter galaxies (SMGs), which is about 200~Myr \citep[e.g.,][]{Narayanan:2010fk}.
The star formation history and high stellar mass (about ten times higher than the median stellar mass in typical $z\sim3$
LBG sample \citep{2001ApJ...562...95S}) in J1432+3358 indicate that it is a massive system with a long-term intensive 
star formation process rather than a low massive system which harbors a star formation burst.
However, the physical properties derived from SED fitting are not reliable due to the weak constraints on the star formation time scale ($\tau$).
The star formation ages $t_{sf}$ have significant degeneracies with the $\tau$ values \citep[see details in][]{2001ApJ...562...95S,Shapley:2005lr},  
which could affect our above conclusion.  
On the other hand, the stellar mass estimation is much more reliable, because the 
rest-frame NIR M/L is reasonably constraint by the SED fitting and nearly independent on the $\tau$ values.

The gas mass can be estimated using the global Kennicutt-Schmidt law \citep{1998ApJ...498..541K}:
\begin{equation}
\Sigma_{\rm gas} = 361\times\left(\frac{\Sigma_{\rm SFR}}{1~M_{\sun}~{\rm yr}^{-1}~{\rm kpc}^{-2}}\right)^{0.71} ~M_{\sun}~{\rm kpc}^{-2},
\end{equation}
where $\Sigma_{\rm SFR} = {\rm SFR}/r_e^2$ and the gas mass, M$_{\rm gas}$, can be derived from 
 $\Sigma_{\rm gas}\times r_e^2$, which is $3.7\times10^{10}$~M$_{\sun}$. The fraction of gas
 in J1432+3358, $M_{\rm gas}/(M_{\rm gas}+M_{\rm stellar})$, is 0.2. The low gas fraction is consistent 
 with the gas fraction in the UV-selected galaxies at the most massive end \citep{2006ApJ...644..813E}
 and comparable to that in sub-millimeter galaxies \citep[]{2008ApJ...680..246T}.

\section{Discussion}
J1432+3358 shows unique properties compared to normal LBGs characterized by: 
(1) a high SFR (about 300~M$_\sun$~yr$^{-1}$),
(2) a 3:1 merger-like morphology with a 1.0$''$ separation of the two components,
(3) a high stellar mass ($1.3\times10^{11}$~M$_{\sun}$), and
(4) a long continuous star formation history (630~Myr), which is twice longer
than median star formation history of LBGs.

One of the key questions is regarding the nature of UV ultra-luminous 
LBGs. By integrating the UV luminosity function \citep[e.g.,][]{2009ApJ...692..778R},
we find that the space number density of the UV ultra-luminous LBGs with $L>7L^*$ 
is a few $10^{-7}~$ Mpc$^{-3}$ in the redshift range from 2.7 to 3.3, 
which corresponds to a spatial number density
of $\sim0.3$~deg$^{-2}$. 
This is consistent with 
the result that there is only one UV ultra-luminous LBG found in the 9~deg$^2$ {\bootes}
within uncertainties. 
The space number density of UV ultra-luminous LBGs is 
about 2-3 orders of magnitude smaller than that of typical star-forming galaxies 
(e.g., LBGs, Bzk galaxies) at $z\sim2-3$, which is 
about $10^{-4}-10^{-5}$~Mpc$^{-3}$ 
\citep[e.g.,][]{2005ApJ...633..748R}. 
On the other hand, the space density of galaxy with its stellar mass greater than 
10$^{11}$~M$_{\sun}$ is about $10^{-4}$~Mpc$^{-3}$
 based on the stellar mass function \citep[e.g.,][]{2005ApJ...619L.131D}.  
The low space density of UV ultra-luminous LBGs can be interpreted
 by the following two different scenarios: the first one is that UV ultra-luminous 
 LBGs can only be found in a short evolutionary phase for the most intensive 
 star-forming galaxies. Though the SED-fitting result shows that 
the star formation time ($t_{sf}$) in J1432+3358 is 630~Myr, this galaxy
may only be selected by UV-selection method during a short time period, i.e., 
most of the time
the galaxy is highly obscured by dust.  
The other interpretation is that most of the galaxies that form stars at high intensity
are dusty and highly obscured, e.g., SMGs \citep[][]{2005ApJ...622..772C} or DOGs (dust-obscured galaxies)
\citep[][]{Dey:2008qy,Fiore:2008fr},
and only small fraction of galaxies that show high SFR is unobscured by dust.

Another crucial question is whether the extremely high SFR in J1432+3358 is triggered by
galaxy major mergers or fueled by rapid accretion of 
cold gas from the intergalactic medium (IGM). 
\citet{Hopkins:2008fj} investigate the role of mergers in the evolution of starburst and quasars
and suggest an evolution track from a `typical'  galaxy to a gas-rich major merger 
\citep[see a schematic outline
in Figure~1 in][]{Hopkins:2008fj}. In this scenario, at  the early stage of merger, i.e. their phase (c),
the two interacting galaxies are within one halo but still well separated and can be identified as 
a merger pair \citep[e.g.,][]{Lotz:2004kx}. 
In this stage, the SFR starts to increase to
about 100 $M_{\sun}$~yr$^{-1}$ due to the tidal torques, but the enhanced effect of the SFR
is relatively weak compared to the latter coalescence phase.
The timescale of this phase is about several million years.
The AGN activity in this phase is relatively low. These features are consistent with the properties of 
J1432+3358.  
 
A tight correlation between the stellar mass ($M_{\rm stellar}$)
and SFR (main sequence) in normal star forming galaxies has been 
found in both local and high-redshift universe 
\citep[e.g.,][]{2007ApJ...670..156D}. 
The mean value of the specific
SFR (sSFR=SFR/$M_{\rm stellar}$) is about $1.8\times10^{-9}$~yr$^{-1}$
in star-forming galaxies at $z\sim2$ at the high stellar mass end ($10^{11.0}$~M$_{\sun}$$<M_{\rm stellar}< $$10^{11.5}$~M$_{\sun}$). The galaxies with sSFR greater than $\sim5.6\times10^{-9}$~yr$^{-1}$
are considered to be off the main sequence and suggested to be merger-driven starburst galaxies \citep[e.g.,][]{Rodighiero:2011fk}. 
The sSFR of J1432+3358 is 
$1.9\times10^{-9}$~yr$^{-1}$, which would locate our galaxy right on the galaxy 
main sequence at $z\sim2$. 
This result implies that the SFR in J1432+3359 is not enhanced by the merger process,
which is also consistent that this system is in the early phase of the merger process.
However, above conclusion is based on the SED
fitting results which are subject to large systematic uncertainty
as discussed in Section~3.3.
Therefore, further high resolution space-based imaging
and ground-based IFU observations will be requested
to provide more firm evidence 
to distinguish whether J1432+3358 is
a major merger or a clumpy disk galaxy.

The outflow properties of J1432+3358 can be studied and compared to the typical 
LBGs. 
\citet{2010ApJ...717..289S} does not find a correlation between the SFR and the velocity
of the wind in the UV-selected galaxies at $z\sim2$, which is found in other
high-redshift galaxies \citep{2009ApJ...692..187W}.  
One interpretation 
is that the SFR dynamic range in \citet{2010ApJ...717..289S} sample is too small
to reveal the relation.
With the SFR an order of magnitude higher than typical $z\sim2$ UV-selected galaxies,
J1432+3358 will provides enough dynamic range to check wether there exists correlation between the SFR and outflow velocity. 
Comparing J1432+3358 to the typical $z\sim2$ UV-selected galaxies \citep{2010ApJ...717..289S},
we do find that the outflow velocity increases with SFR, and roughly follows the
relation found in \citet{2009ApJ...692..187W}, $v_{out}\propto SFR^{0.3}$.


\acknowledgments
We would like to thank the anonymous referee for providing constructive comments and 
help in improving the manuscript.
This work made use of images and/or data products provided by the NOAO Deep Wide-Field Survey ((Jannuzi and Dey 1999), 
which is supported by the National Optical Astronomy Observatory (NOAO)). NOAO is operated by AURA, Inc., 
under a cooperative agreement with the National Science Foundation.
We thank the LBTO, NOAO, and Gemini staff for their great support in preparing the 
observing and carrying out the observing.  
FB thanks M. Kriek for providing her IDL SED-fitting package and C. Peng for
discussion on GALFIT. YTL is grateful to the HyperSuprimeCam software development team for
the aid of reduction of the SuprimeCam data.
FB, XF, LJ and IM acknowledge support from a Packard Fellowship for Science and 
Engineering and NSF grant AST 08-06861. 



{\it Facilities:} \facility{LBT}, \facility{MMT}, \facility{Gemini-N},\facility{KPNO 4m Mayall},\facility{Spitzer},\facility{Subaru}

\bibliography{LBG_paper,paper}

\clearpage




\end{document}